\documentclass[12pt]{article}
\usepackage[]{epsfig}
\usepackage[centertags]{amsmath}
\usepackage{amsfonts}
\usepackage{amssymb}
\usepackage[english]{babel}
\usepackage{amsmath, amsthm, slashed,url}
\usepackage{color}
\begin{document}

\title{\bf Note on generalization of Jackiw-Pi Vortices}
\author{ 
Lucas Sourrouille
\\
{\normalsize\it  Universidad Nacional Arturo Jauretche, 1888, 
}\\ {\normalsize\it Florencio Varela, Buenos Aires, Argentina}
\\
{\footnotesize sourrou@df.uba.ar } } \maketitle

\abstract{We analyze two abelian Higgs systems with nonstandard kinetic terms. First, we consider a model involving the Maxwell 
term. For a particular choice of the nonstandard kinetics, we are able to obtain generalized Jackiw-Pi vortices. We, also, 
analyze a second model which is a generalization of the Jackiw-Pi model. In this case we show that the system support the 
Nielsen-Olesen vortices as 
solutions. }

\vspace{0.5cm}



\vspace{1cm}
\section{Introduction}

The two dimensional abelian Higgs model coupled to gauge fields, whose dynamics is dictated by a Maxwell term, admits topological 
vortex solutions \cite{1,ph,gl,no}. Furthermore, with a specific choice of coupling constants, the model presents particular 
properties such as supersymmetry extension \cite{WO}-\cite{WO2} and the reduction of the field equations to first order 
differential 
equations, often called Bogomolnyi equations \cite{bogo}. In addition Chern-Simons Higgs theory in (2+1)-dimensions without 
Maxwell term presents Bogomolnyi equations for a sixth order potential \cite{JW, JW1}. In this model there exist both 
topological and 
non-topological solitons.  These solitons carry not only magnetic flux, but also electric charge.

In the recent years, theories with nonstandard kinetic term, named $k$-field models, have received much attention. 
The $k$-field models are mainly in connection with effective cosmological models\cite{BMV,APDM} as well as the strong interaction 
physics, strong gravitational waves and dark matter. One interesting aspect to analyze in these models concern to its topological 
structure. 
In the context of soliton solution, 
several works have be done showing that the $k$-theories can support 
topological soliton solutions both in models of matter as in gauged models\cite{SG}-\cite{SG3} and \cite{ls}-\cite{ls7}. These 
solitons have certain features 
such as their characteristic size, which are not necessarily those of the standard models\cite{B}. Other interesting aspects are 
that they do not interact at large distances and they are, in general, not self-dual.

In this note we are interested in studying a generalization of the abelian Maxwell-Higgs model, by introducing a nonstandard 
kinetic terms in the Lagrangian. We introduce the nonstandard dynamics by a function $\omega$, which depends on the Higgs field.
We study the Bogomolnyi limit for such system. We are able to obtain, for a particular choice of the $\omega$ and the potential 
term, a generalization of the Bogomolnyi equations corresponding to a nonrelativistic Chern-Simons-Higgs theory usually know as 
Jackiw-Pi model \cite{JP, JP1}. We construct the solutions of theses equations and we show that the Jackiw-Pi vortices may be 
obtained 
as a particular case. As particular feature, we show that our vortex solutions have no electric charge. This is a difference with 
the solutions of the Jackiw-Pi model.   

We further propose a generalization of the Jackiw-Pi model. In particular, will show that choosing a suitable $\omega$, the 
Bogomolnyi equations of the Maxwell-Higgs model are obtained. The soliton solutions of these equation are identical in form to 
the Nielsen-Olesen vortices. Nevertheless, our vortex solution have electric charge, a feature which is not present in the usual 
Nielsen-Olesen vortices.

\section{Generalized Jackiw-Pi vortices from abelian Maxwell-Higgs model}
\label{2}

Following the the works cited in Ref.\cite{ls}-\cite{ls7}, we
start by considering a generalized $(2+1)$-dimensional Maxwell
model coupled to bosonic field $\phi$.
The dynamics of this model is descried by the action
\begin{eqnarray}
S = \int \,\, d^3 x \;\;\ \Big(-\frac{1}{4}F_{\mu \nu}F^{\mu \nu}
 +\omega (\rho) [\frac{1}{2}|D_0 \phi|^2
-\frac{1}{2}|D_i \phi|^2]- V(\rho)  \Big)
\label{Acg1}
\end{eqnarray}
The covariant derivative, here, is defined as
\begin{eqnarray}
D_{\mu}= \partial_{\mu} + ieA_{\mu} \;\;\;(\mu =0,1,2)
\end{eqnarray}
The metric tensor is  $g^{\mu \nu}=(1,-1,-1)$ and $\epsilon^{\mu\nu\lambda}$
is the totally antisymmetric tensor such that $\epsilon^{012}=1$.
\\
Notice, that we have replaced the usual kinetic term $|D_i \phi|^2$ by a more generalized term 
$\omega (\rho)|D_i \phi|^2$, where $\omega(\rho)$ is a 
dimensionless functions of the complex scalar field $\phi$. Here, 
$V(\rho)$ is the scalar field potential to be determined below, being $\rho=\phi^\dagger \phi$.

Variation of this action yields the field equations.
\begin{eqnarray}
&&\frac{\partial\omega(\rho)}{\partial \phi^*}|D_0 \phi|^2- \frac{\partial\omega(\rho)}{\partial \phi^*}|D_i 
\phi|^2 + \omega(\rho) D_0 D^0 \phi - \omega(\rho) D_i D^i \phi -2\frac{\partial V}{\partial \phi^*} =0
\nonumber \\[3mm]
&&\partial^\nu[F_{i \nu}] = -e j_i \omega(\rho)
\nonumber \\[3mm]
&&\partial^\nu[F_{0 \nu}] = -e j_0 \omega(\rho)
\label{geEq}
\end{eqnarray}
where $j_0 = \frac{1}{2i}\Big(\phi^* D_0 \phi - (D_0 \phi)^* \phi \Big)$  and $j_i = \frac{1}{2i}\Big(\phi^* D_i \phi - 
(D_i \phi)^* \phi \Big)$. 
From the last equation of (\ref{geEq}), we see that $A_0=0$ gauge may be chosen in consitent way for static solutions. 
\\
Here, we are interested in time-independent soliton solutions that ensure the finiteness of the action (\ref{Acg1}).
These are the stationary points of the energy which for the static field configuration reads
\begin{eqnarray}
E = \int \,\, d^2 x \;\;\ \Big(\frac{1}{2}B^2
+\frac{1}{2}\omega (\rho)|D_i \phi|^2 + V(\rho)  \Big)
\label{AcgM1}
\end{eqnarray}
In order to find self-dual soliton solutions, we consider the following choice for the function 
$\omega(\rho)$  
\begin{eqnarray}
\omega(\rho) = (n+1)\rho^{n} \;,
\label{omega}
\end{eqnarray}
where $n$ is a real number that satisfies $n\geq 0$. 
\\
To proceed, we need the fundamental identity 
\begin{eqnarray}
|D_i \phi|^2 = |( D_1 \pm iD_2)\phi|^2 \mp eB|\phi|^2 \pm \epsilon^{ik} \partial_i j_k \;,
\label{iden}
\end{eqnarray}
where $i$ and $k$ may take the values $1$ and $2$.
Then, we may rewrite the energy (\ref{AcgM1}) as
\begin{eqnarray}
E = \int \,\, d^2 x &\Big(& \frac{1}{2}B^2 + \frac{(n+1)}{2}\rho^n|( D_1 \pm iD_2)\phi|^2 \mp 
\frac{(n+1)}{2}\rho^{n+1} eB 
\nonumber\\
&+& V(\rho) \pm  \frac{(n+1)}{2}\rho^n \epsilon^{ik} \partial_i j_k\Big)
\label{EJP2}
\end{eqnarray}
The last term of this integral may be integrated by  parts. Indeed, after a bit of algebra,  it is not difficult to show that
\begin{eqnarray}
\int \,\, d^2 x \;\; 
\epsilon^{ik}\rho^n \partial_i j_k = \int \,\, d^2 x \;\; e\Big( \epsilon^{ik} \rho^n A_k \partial_i \rho + \epsilon^{ik} 
\rho^n 
\partial_i A_k \rho \Big)\;,
\label{}
\end{eqnarray}
which may be rewritten in a more suitable form,
\begin{eqnarray}
\int \,\, d^2 x \;\; 
\epsilon^{ik}\rho^n \partial_i j_k = \int \,\, d^2 x \;\; e\Big(\frac{\epsilon^{ik}}{n+1} A_k \partial_i \rho^{n+1} + 
\epsilon^{ik} \rho^{n+1} \partial_i A_k \Big)
\label{a}
\end{eqnarray}
Again, integrating by parts, we have, 
\begin{eqnarray}
\int \,\, d^2 x \;\; 
\epsilon^{ik}\rho^n \partial_i j_k = e\int \,\, d^2 x \;\; 
\epsilon^{ik}\partial_i A_k \rho^{n+1} (-\frac{1}{n+1} +1) = \frac{ne}{n+1}\int \,\, d^2 x \;\; B \rho^{n+1}
\label{a1}
\end{eqnarray}
Thus, using (\ref{a1}), the energy (\ref{EJP2}) is rewritten as
\begin{eqnarray}
E = \int \,\, d^2 x \;\;\ \Big(
\frac{(n+1)}{2}\rho^n|( D_1 \pm iD_2)\phi|^2 \mp 
\frac{e}{2}\rho^{n+1} 
B + V(\rho)+\frac{1}{2}B^2 \Big)
\label{AcgM2}
\end{eqnarray}
In the case that we choose 
\begin{eqnarray}
V(\rho) = \frac{e^2}{8} \rho^{2(n+1)}\;,
\label{p1}
\end{eqnarray}
we can write the last two terms of (\ref{AcgM2}) as 
\begin{eqnarray}
V(\rho)+\frac{1}{2}B^2 = \frac{1}{2}[B \mp \frac{e}{2}\rho^{(n+1)}]^2 \pm \frac{e}{2}B\rho^{(n+1)}
\end{eqnarray}
Then,
\begin{eqnarray}
E = \int \,\, d^2 x \;\;\ \Big(\frac{1}{2}[B \mp \frac{e}{2}\rho^{(n+1)}]^2 +
\frac{(n+1)}{2}\rho^n|( D_1 \pm iD_2)\phi|^2 \Big)
\label{AcgM3}
\end{eqnarray}
The energy is bounded below by zero and the bound is saturated by the fields satisfying the first-order equations,
\begin{eqnarray}
&&\phi^n(D_1 \pm iD_2)\phi =0
\nonumber \\[3mm]
&&B =  \pm \frac{e}{2} \rho^{(n+1)}
\label{bog}
\end{eqnarray}
Here, it is interesting to analyze these equations when $n=0$. In that case the equations (\ref{bog}) becomes,
\begin{eqnarray}
&&(D_1 \pm iD_2)\phi =0
\nonumber \\[3mm]
&&B =  \pm \frac{e}{2} \rho
\label{bogo2}
\end{eqnarray}
which may be compared with the self-duality equations of the Jackiw-Pi model \cite{JP}. Indeed, if we choose the plus sign in 
the 
second equation of (\ref{bogo2}) we arrive to the Bogomolnyi equations of the Jackiw-Pi model,
\begin{eqnarray}
&&(D_1 \pm iD_2)\phi =0
\nonumber \\[3mm]
&&B = \frac{ e}{2} \rho
\label{bogo3}
\end{eqnarray}
Also. if $n=0$, the potential term (\ref{p1}), becomes a $\phi^4$ potential as in the Jackiw-Pi model.
To solve the Bogomolnyi equations of the Jackiw-Pi model is usual decompose the scalar field $\phi$
into its phase and magnitude:
\begin{eqnarray}
\phi = \rho^{\frac{1}{2}} e^{i\alpha}
\label{}
\end{eqnarray}
Then the first of the self-duality equations (\ref{bogo3}) determines the gauge field
\begin{eqnarray}
A_i = \frac{1}{2ie \rho}\Big(\mp i \epsilon^{ij} \partial_j \rho -\phi^{\dagger} \partial_i \phi + \phi \partial_i \phi^\dagger 
\Big)
\label{a11}
\end{eqnarray}
everywhere away from the zeros of the scalar field. Thus, using (\ref{a11}) the second self-duality equation in (\ref{bogo3}) 
reduces to a nonlinear 
elliptic equation for the scalar field density $\rho$,
\begin{eqnarray}
\pm e^2\rho  = \nabla^2 \log \rho 
\label{}
\end{eqnarray}
This elliptic equation, known as the Liouville equation, is exactly solvable,
\begin{eqnarray}
\rho = \frac{2}{e^2} \nabla^2 \log \Big(1 + |f|^2\Big)
\label{}
\end{eqnarray}
where $f=f(z)$ is a holomorphic function of $z= x_1 +ix_2$.
General radially symmetric solutions may be obtained by taking $f(z)=\Big(\frac{z_0}{z}\Big)^N$. Then, we have 
\begin{eqnarray}
\rho = \frac{8 N^2}{e^2 r_0^2} \frac{ \Big(\frac{r}{r_0}\Big)^{2(N-1)}}{\Big(1+ (\frac{r}{r_0})^{2N} \Big)^2}
\label{}
\end{eqnarray}
This vanish as $r\to\infty$ and is nonsingular at the origin for $|N|\geq 0$ but for  $|N|> 0$, the vector 
potential behaves as $A_i (r) \sim - \partial_i \alpha \mp 2(N-1)\epsilon_{ij}\frac{x^j}{r^2}$. Therefore we can avoid
singularities in the the vector potential at the origin if we choose the phase of $\phi$ to be $\alpha = \pm \theta(N-1)$. Then,
the self-dual $\phi$ field is
\begin{eqnarray}
\phi = \frac{\sqrt{2}2 N }{e r_0} \Big(\frac{(\frac{r}{r_0})^{N-1}}{1+ (\frac{r}{r_0})^{2N}} \Big) e^{\pm 
2i(N-1)\theta}
\label{}
\end{eqnarray}
Requiring that $\phi$ be single-valued we find that $N$ must be an integer, and for $\rho$ to decay at infinity
we require that $N$ be positive.
\\
This procedure may be used to solve the equations (\ref{bog}).
Indeed, we can define a new field $\psi=\phi^{n+1}$. Thus, it is not difficult to show that the equations (\ref{bog}) may be 
rewritten, in terms of $\psi$, as 
\begin{eqnarray}
&&(D_1^{'} \pm iD_2^{'})\psi =0
\nonumber \\[3mm]
&&B =  \pm \frac{e}{2}(\psi^\dagger \psi)
\label{bogo5}
\end{eqnarray}
where $D_i^{'}= \frac{1}{n+1}\partial_i +ieA_i$. So, the set of the equations (\ref{bogo5}) is basically the same as 
(\ref{bogo2}), and therefore it has the same solution, i.e.
\begin{eqnarray}
\psi = \frac{\sqrt{2}2 N }{\sqrt{e^2(n+1)}r_0} \Big(\frac{(\frac{r}{r_0})^{N-1}}{1+ (\frac{r}{r_0})^{2N}} \Big) e^{\pm 
2i(N-1)\theta}
\label{psi1}
\end{eqnarray}
Notice that in the denominator of (\ref{psi1}) appears the factor $(n+1)$. 
We can rewrite (\ref{psi1}) in a more compact form
\begin{eqnarray}
\psi = g(r) e^{\pm 
i(N-1)\theta}
\label{}
\end{eqnarray}
Then, the field $\phi$ is 
\begin{eqnarray}
\phi = (g(r))^{\frac{1}{n+1}} e^{\pm 
im\theta}
\label{}
\end{eqnarray}
being $m = \frac{2(N-1)}{n+1}$.Since, $\phi$ must be single-valued we require that $\frac{2(N-1)}{n+1}$ be an integer.
\\
It is interesting to note that our solutions are electrically neutral, i.e. $j_0=0$ for our solutions. This constitute a 
difference with the Jackiw-Pi vortices, since it is well know that these vortices are electrically charged.  Thus, in the 
particular 
case $n=0$, we have that our solutions are mathematically identical to the Jackiw-Pi vortices, although from the physical point 
of view our vortices do not have electric charge. 
\\
Another interesting point refers to the possible self-dual question obtained from the model described by the action 
(\ref{Acg1}). 
Indeed, we can obtain another set of self-dual equations if the self-dual potential is chosen to be 
\begin{eqnarray}
V(\rho) = \frac{e^2}{8} (1- \rho^{(n+1)})^2\;,
\label{p21}
\end{eqnarray}
In this case it is not difficult to see that the energy (\ref{AcgM2}) may be rewritten as
\begin{eqnarray}
E = \int \,\, d^2 x \;\;\ \Big(\frac{1}{2}[B \pm \frac{e}{2}(1-\rho^{(n+1)}]^2 \mp \frac{e}{2} B +
\frac{(n+1)}{2}\rho^n|( D_1 \pm iD_2)\phi|^2 \Big)
\label{}
\end{eqnarray}
We see that the energy is bounded below by a multiple of the magnitude
of the magnetic flux (for positive flux we choose the lower signs, and for negative flux we choose
the upper signs):
\begin{eqnarray}
E \geq \frac{e}{2} |\Phi|
\end{eqnarray}
In order to the energy be finite the covariant derivative must vanish asymptotically. This fixes the  behavior of the gauge field 
$A_i$ and implies a non-vanishing magnetic flux:
\begin{eqnarray}
\Phi = \int \,\,d^2 x B = \oint_{|x|=\infty} \,\, A_i dx^i  = 2\pi N
\end{eqnarray}
where $N$ is a topological invariant which takes only integer values.
The bound is saturated by fields satisfying the first-order self-duality equations:
\begin{eqnarray}
& & \phi^n( D_1 \pm iD_2)\phi =0
\nonumber\\
& &
B= \mp \frac{e}{2} (1-\rho^{(n+1)})
\label{w2}
\end{eqnarray}
This constituted a new set of self-duality equations for the action (\ref{Acg1}) of the paper. In the particular case of $n=0$ 
equations 
(\ref{w2}), becomes the well know equations of the abelian Higgs model  \cite{no}.

\section{Nielsen-Olesen vortices from a generalized Jackiw-Pi model}
\label{3}
Suppose that instead of action (\ref{Acg1}) we had a generalization of the Jackiw-Pi model
\begin{eqnarray}
S = \int \,\, d^3 x & \Big( \frac{\kappa}{2}\epsilon^{\mu \nu \rho} A_\mu \partial_\nu A_\rho
+i\omega(\rho)\phi^* D_0 \phi - \frac{1}{2m} |D_i \phi|^2 - V(\rho) \Big)
\label{Ac3}
\end{eqnarray}
As in model, (\ref{Acg1}) $\omega(\rho)$ is positive-definite 
dimensionless functions of the complex scalar field $\phi$.
\\
The equations of motion for this system are given by
\begin{eqnarray}
&&i\Big(\frac{\partial\omega(\rho)}{\partial \phi^*}\phi^*D_0 \phi + \omega(\rho)D_0 \phi\Big) = - \frac{1}{2m} D_i^2 \phi +
\frac{\partial V(\rho)}{\partial \phi^*}
\nonumber \\[3mm]
&&B=\frac{e}{\kappa} \omega(\rho)\rho
\nonumber \\[3mm]
&&E^i = -\frac{1}{\kappa} \epsilon^{i j} j_i
\label{EqM3}
\end{eqnarray}
where the second equation is the Gauss law. 
\\
The theory may be descried in terms of the Hamiltonian formulation as
\begin{eqnarray}
H = \int \,\, d^2 x & \Big(  \frac{1}{2m} |D_i \phi|^2 + V(\rho) \Big)
\label{}
\end{eqnarray}
which may be rewritten using the Gauss Law and the identity (\ref{iden}) in the form
\begin{eqnarray}
E = \int \,\, d^2 x & \Big(  \frac{1}{2m} |D_{\pm} \phi|^2 \mp \frac{e^2}{2m\kappa} \omega(\rho)\rho^2 + V(\rho) \Big)
\label{E2x}
\end{eqnarray}
Here, it will be interesting to consider the following $\omega(\rho)$ function,
\begin{eqnarray}
\omega(\rho) = \pm \kappa (\rho -1)\rho^{-1}
\label{omega0}
\end{eqnarray}
Thus, we have two possible cases: 
\begin{eqnarray}
\omega(\rho) =  \kappa (\rho -1)\rho^{-1} \;\;\;,
\omega(\rho) = - \kappa (\rho -1)\rho^{-1}
\end{eqnarray}
The first case correspond to the upper signs in the energy (\ref{E2x}), the second to the lower signs.
\\
Then, we can introduce (\ref{omega0}) into (\ref{E2x}) to obtain 
\begin{eqnarray}
E = \int \,\, d^2 x & \Big(  \frac{1}{2m} |D_{\pm} \phi|^2 \mp \frac{e^2}{2m} (\rho -1)\rho + V(\rho) \Big)\;,
\label{E2x1}
\end{eqnarray}
For $\omega(\rho)$ taking the form of the formula (\ref{omega0}), the Gauss law of the equation (\ref{EqM3}) becomes
\begin{eqnarray}
B=\frac{e}{\kappa}\omega(\rho)\rho= \pm e (\rho -1)
\label{Gauss1}
\end{eqnarray}
Now, suppose that we choose a symmetry breaking $\phi^4$ potential as is usual in Maxwell-Higgs model
\begin{eqnarray}
V(\rho) = \frac{e^2}{2m} (\rho-1)^2
\label{pot}
\end{eqnarray}
Then, the energy (\ref{E2x1}) becomes,
\begin{eqnarray}
E = \int \,\, d^2 x & \Big(  \frac{1}{2m} |D_{\pm} \phi|^2 \mp \frac{e}{2m} B \Big)\;,
\label{E2x2}
\end{eqnarray}
Here, the situation is very similar to the case of Nielsen-Olesen and Chern-Simons-Higgs vortices, in the sense that these 
models are bounded bellow by a multiple of magnitude of the magnetic flux (for positive flux we choose the lower signs, and for 
negative flux we choose
the upper signs):
\begin{eqnarray}
E \geq \frac{e}{2m} |\Phi|
\end{eqnarray}
The requirement of finiteness of the energy fix the magnetic flux. This requirement imply that the covariant derivative must 
vanish asymptotically, which establishes the behavior of the gauge field $A_i$,
\begin{eqnarray}
\Phi = \int \,\,d^2 x B = \oint_{|x|=\infty} \,\, A_i dx^i  = 2\pi N
\end{eqnarray}
where $N$ is an integer and a topological invariant. Then, the topological bound is saturated by fields satisfying the 
first-order 
self-duality equations
\begin{eqnarray}
& &D_{\pm}\phi = ( D_1 \pm iD_2)\phi =0
\\
& &
B=\pm e (\rho -1)
\end{eqnarray}
Here, we have a couple of equations which are identical to the self-duality equations of the abelian Maxwell-Higgs model 
\cite{bogo}. Nevertheless, our vortices are physically different from the Nielsen-Olesen solutions \cite{no}. Indeed, our 
vortices not only carry magnetic flux, but also $U(1)$ charge. From the Gauss law (\ref{Gauss1}), we know 
$B=\frac{e}{\kappa}\omega(\rho)\rho$ and from the Noether theorem we have that the conserved charge associated to $U(1)$ 
transformation
\begin{eqnarray}
\delta \phi = -ie \phi \;,
\end{eqnarray}
is 
\begin{eqnarray}
Q_{U(1)} = \int \,\, d^2 x \;\; j^0 =  \int \,\, d^2 x \frac{\partial \cal{L}}{\partial(\partial_0 \phi)}\delta \phi =
e\int \,\, d^2 x \;\; \omega(\rho)\rho
\end{eqnarray}
Thus, the magnetic flux is proportional to the $U(1)$ charge,
\begin{eqnarray}
\Phi = \frac{1}{\kappa} Q_{U(1)} \;,
\end{eqnarray}
which is a particularity of the Chern-Simons models \cite{JW}.
\\
So, starting from a generalization of the Jackiw-Pi model we were able to obtain the self-duality equations of the Maxwell-Higgs 
model. However, our vortices are different from the Nielsen-Olesen vortices. The difference lies in the fact that, here, our 
vortices not only carry magnetic flux, as in the Higgs model, but also $U(1)$ 
charge.

\vspace{0.6cm}

{\bf Acknowledgements}
\\

This work is supported by CONICET.

\end{document}